\documentclass[12pt,preprint]{aastex}

\shorttitle{A high AGN fraction in red cluster galaxies}
\shortauthors{Martini \etal}

\slugcomment{Accepted for publication in ApJL}

\newcommand{\eg}{{\rm e.g.}}
\newcommand{\etal}{{\rm et al.}}
\newcommand{\ergs}{erg s$^{-1}$}
\newcommand{\kms}{km s$^{-1}$}
\newcommand{\chandra}{{\it Chandra}}

\begin{document}

\title{An unexpectedly high AGN fraction in red cluster galaxies}

\author{Paul Martini, Daniel D. Kelson, John S. Mulchaey, and 
Scott C. Trager\altaffilmark{1}}

\altaffiltext{1}{Hubble Fellow}

\affil{Carnegie Observatories, 813 Santa Barbara St., Pasadena, CA 91101-1292 \\
martini@ociw.edu, kelson@ociw.edu, mulchaey@ociw.edu, sctrager@ociw.edu}

\begin{abstract}

As part of a program to study the evolution of active galactic nuclei (AGN) 
in clusters of galaxies, we present our results for Abell~2104. 
A deep \chandra\ observation of this massive, $z = 0.154$ cluster reveals a 
significant X-ray point source excess over the expectations of blank fields, 
including eight X-ray counterparts with $R<20$ mag. 
Our spectroscopy shows that all six X-ray sources associated with red 
counterparts are cluster members and their X-ray properties are 
consistent with all of them being AGN. Only one of the six has the emission 
lines characteristic of optically selected AGN; the remaining five 
would not have been classified as AGN based on their optical spectra. 
This suggests the existence of a large population of obscured, or at least 
optically unremarkable, AGN in clusters of galaxies. 
These six sources correspond to a lower limit of $\sim 5$\% of the AGN 
fraction in cluster galaxies with $R<20$ mag (rest-frame $M_V = -19.5$ mag) 
and is comparable to the blue galaxy fraction in the cluster.
Such an obscured AGN population in clusters of galaxies has many implications 
for cluster galaxy evolution, the hidden growth of their central, supermassive 
black holes, estimates of the star formation rate at infrared and radio 
wavelengths, and the observed variance in the hard X-ray background. 

\end{abstract}


\section{Introduction}

Both star formation and accretion onto supermassive black holes require 
reservoirs of cold gas. 
The presence of nuclear activity in cluster galaxies is therefore an 
indication of the efficiency with which these galaxies were stripped of 
their cold interstellar medium, and to what extent their central, supermassive 
black holes may continue to grow in cluster environments. 
The evolution of the fraction of cluster galaxies which host active galactic 
nuclei (AGN) is therefore an important component of galaxy evolution in 
clusters. 
Determination of the connection between AGN and the history of infall and 
star formation for host galaxies may also provide interesting constraints on 
the fueling and lifetimes of nuclear activity. 
AGN in clusters of galaxies may be a significant contribution to the radio and 
infrared sources in clusters, and need to be identified 
in order to use such measurements to infer star formation rates 
\citep[\eg][]{duc02}. 

It is not yet known if the AGN fraction in clusters of galaxies shows 
any evolution with redshift, although there is clear evidence for 
evolution in cluster galaxies. 
The fraction of galaxies with recent star formation decreases significantly 
from $z \sim 0.5$ to the present 
\citep[the Butcher-Oemler effect; ][]{butcher78,butcher84} and 
the fraction of poststarburst galaxies also has decreased dramatically 
over this same redshift range \citep{dressler88,dressler99} and much more so 
than in the field \citep{zabludoff96,dressler99}. 
The current best estimate of the mean cluster AGN fraction is $\sim 1$\%
\citep[\eg][]{dressler83,dressler99}, a factor of a few less than the bright 
AGN fraction in the field \citep[\eg][]{huchra92}.
These measurements of the AGN fraction in clusters and the field are
the result of extremely large, spectroscopic surveys of galaxies.
However, the fraction of all galaxies that are bright AGN is small; this makes 
spectroscopic searches inefficient and any evolution in the host galaxy 
population difficult to detect. 
A related problem is that AGN may not be obvious at visible wavelengths, 
whether due to moderate line strengths compared to the host galaxy light, 
or obscuration.

A better way to search for AGN is at X-ray wavelengths. 
While AGN comprise only a small minority of all galaxies at visible 
wavelengths, they are the dominant contribution to the luminous, hard 
(2-10 keV) X-ray point source population. 
We have begun a program to study the AGN content of clusters of galaxies 
that have deep observations with the {\it Chandra X-ray Observatory}. 
As all luminous AGN produce significant hard X-ray emission, we use the 
\chandra\ data to select AGN for spectroscopic observations to determine 
cluster membership. 
\chandra\ data are therefore uniquely suited to identify these AGN due to its 
superior angular resolution compared to all other X-ray satellites. 
In this {\it Letter} we present our discovery of six previously unknown 
AGN in Abell~2104, a rich cluster \citep[Abell class 2;][]{allen92} of 
galaxies at $z = 0.154$ \citep{liang00}.

\section{Observations} 

\begin{figure}
\epsscale{1.0}
\plotone{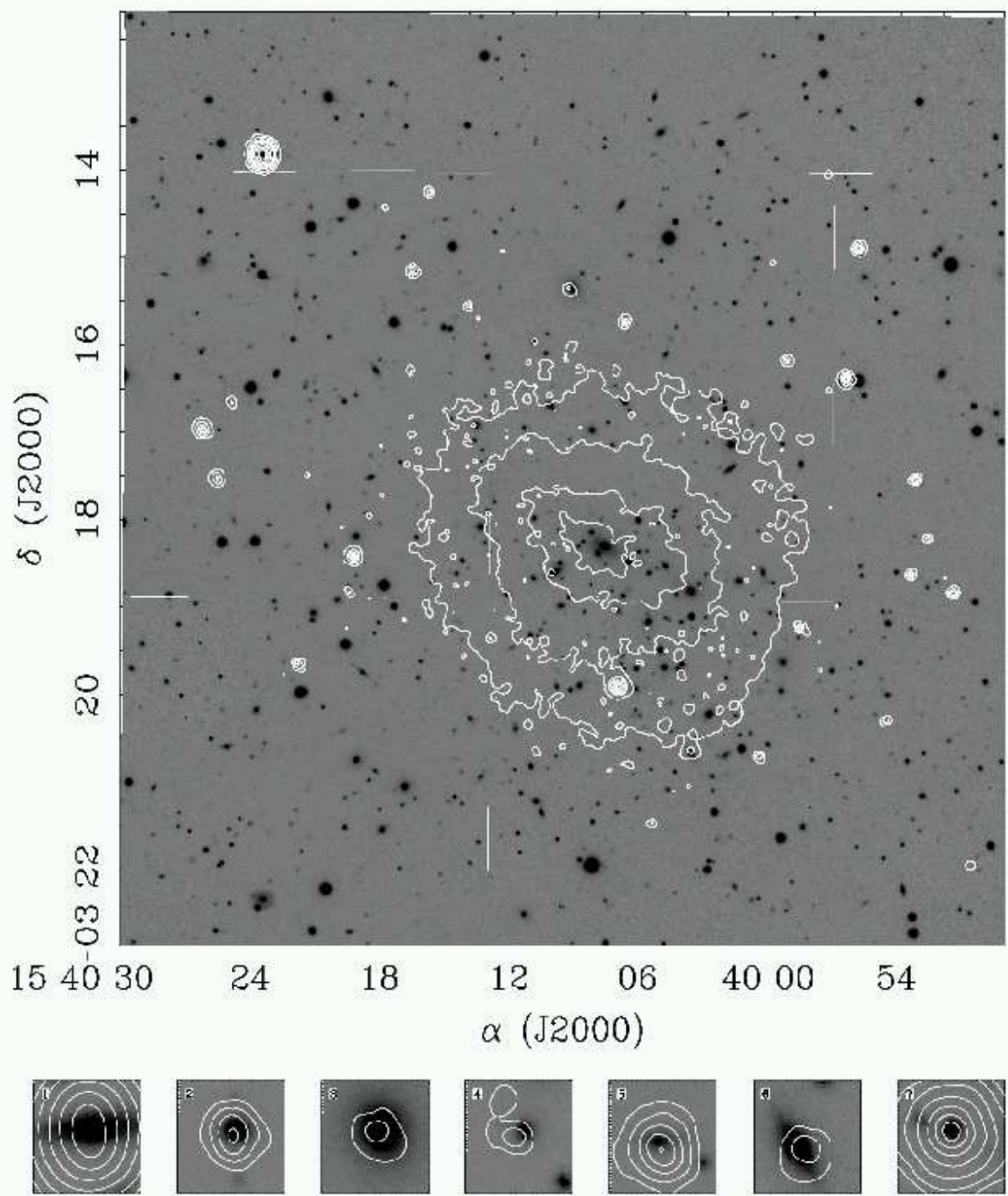}
\caption{\small
A deep, ground-based $R$-band image of the inner $10' \times 10'$ of 
Abell 2104, a cluster of galaxies at $z=0.154$. The X-ray contours 
from the \chandra\ observations are also shown. The lowest contour level 
is $2 \times 10^{-17}$ \ergs cm$^{-2}$ arcsec$^{-2}$ and each contour level 
increases by approximately a factor of two, depending on the detailed 
spectra shape. The individual galaxies shown in the small panels highlight 
the seven X-ray sources from Table~\ref{tbl:obs}. \label{fig:image}}
\end{figure}

\begin{figure}
\epsscale{1.0}
\plotone{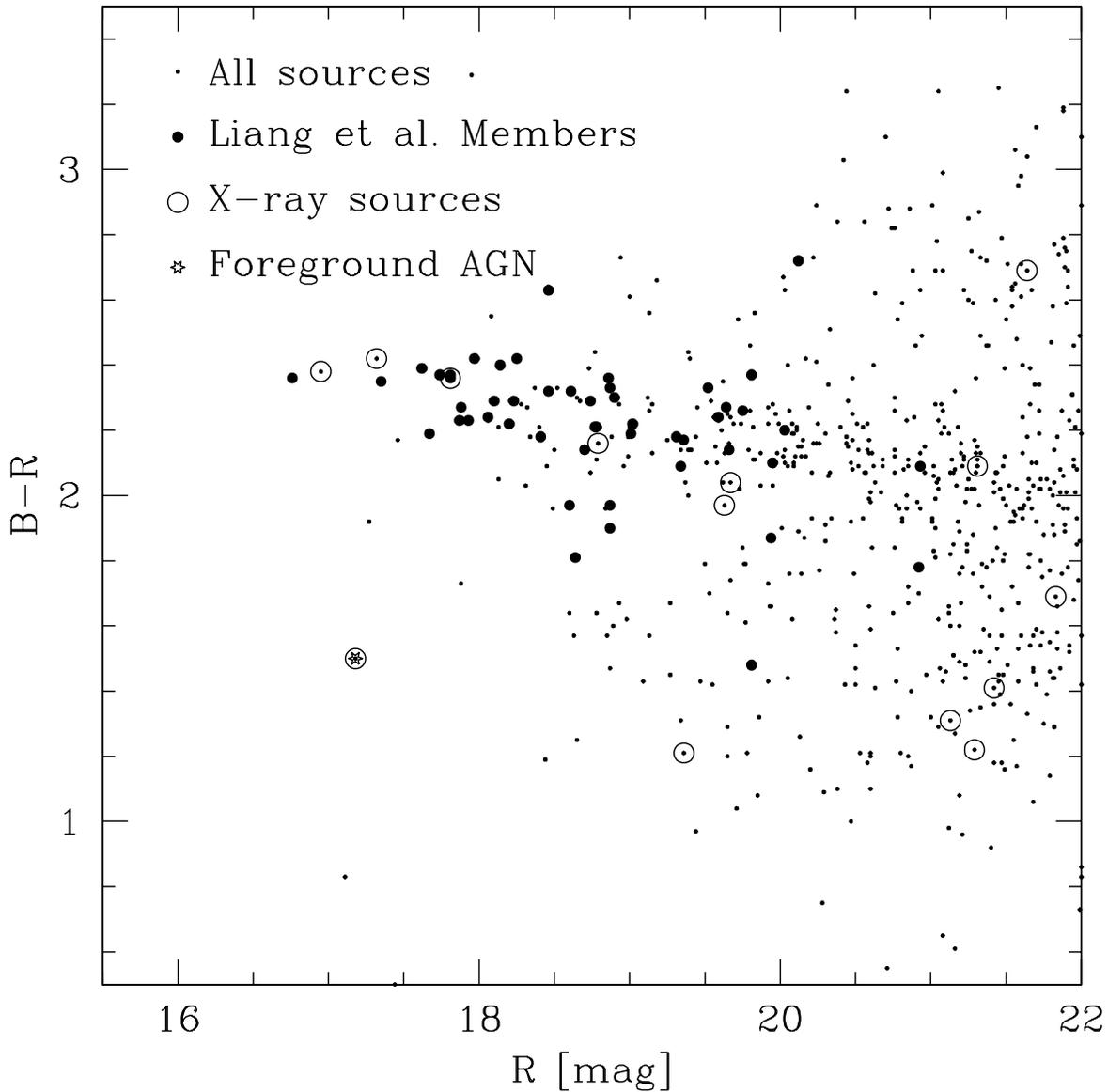}
\caption{\small
$(B-R)-R$ Color--Magnitude diagram for galaxies in our ground-based imaging 
of Abell~2104. The points show the data for all galaxies in the image with
photometric errors $<0.2$ mag. The filled circles show
spectroscopically confirmed cluster members \citep[from][]{liang00},
while the open circles show the counterparts to the 14 Chandra point 
sources with $R<22$ mag. The star marks the foreground AGN.  
There are eight counterparts brighter than $R<20$ mag, while only one or two 
would be expected in a blank field \citep{mushotzky00}. 
We have confirmed that all six, red counterparts are cluster members. 
\label{fig:cmd}}
\end{figure}

Abell~2104 was observed by \chandra\ with ACIS-S on 25 May 2000 for 49.83ks. 
Here we restrict our analysis to the S3 chip, which covers a $8.9'\times8.9'$ 
field of view. The hard X-ray flux limit for this observation is approximately 
$F_X{\rm [2-10keV]} > 10^{-15}$ \ergs cm$^{-2}$. 
We identified point sources in a smoothed version of the 0.8-10 keV image with 
SExtractor \citep{bertin96}. 
Figure~\ref{fig:image} shows the X-ray contours, after subtraction of the 
diffuse cluster component, on a deep 
$R-$band image of the inner $10'\times10'$ of the cluster obtained at the 
2.5m du Pont telescope. 
Fourteen of the X-ray sources have optical counterparts brighter than 
$R<22$ mag; eight have optical counterparts brighter than $R<20$ mag. 
Figure~\ref{fig:cmd} shows a $B-R$ vs.\ $R$ color--magnitude diagram 
({\it small points}) and the spectroscopically confirmed cluster members 
from \citet[][{\it filled circles}]{liang00}.  
The X-ray detections ({\it open circles}) were not observed by \citet{liang00}, 
with the exception of a foreground AGN at $z = 0.0367$ ({\it star} in 
Figure~\ref{fig:cmd}, galaxy \#117 in Liang \etal) and one cluster member 
(our \#6, Liang \etal\ \#314). 

\begin{figure}
\epsscale{1.0}
\plotone{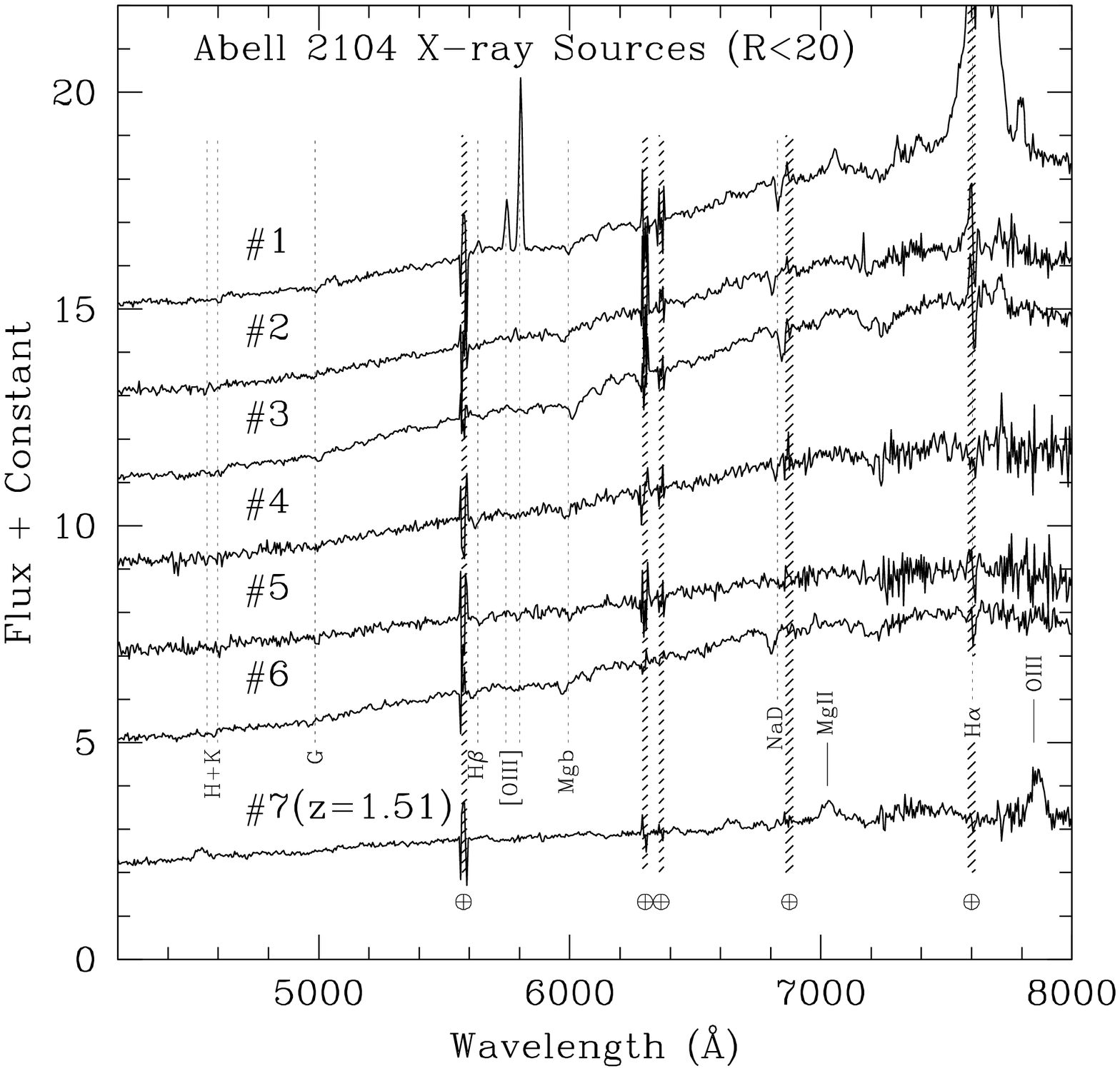}
\caption{\small
Spectra of the seven X-ray sources with $R<20$ mag in the field of Abell 2104.
The dotted lines mark the location of spectral features at the
cluster redshift, while the hatched regions mark the locations of strong 
telluric emission or absorption.
Only one (\#1) of the six galaxies has the strong emission lines 
characteristic of an AGN.
The remaining, bluer object (\#7) is a quasar at $z = 1.51$.
\label{fig:spectra}}
\end{figure}

We observed the seven counterparts with $R < 20$ mag (we excluded the 
foreground AGN) with the 6.5m Baade telescope on 6 April 2002. 
These spectra, shown in Figure~\ref{fig:spectra}, were obtained with the 
Low Dispersion Survey Spectrograph~2 and a $1.25''$ longslit. 
{\it All six} of the red, $R<20$ mag optical counterparts are 
members of Abell~2104. 
The seventh, blue counterpart is a QSO at $z = 1.51$. 
Based on \chandra\ observations of blank fields, we expect one or two 
optically bright X-ray counterparts in an observation with this exposure time
\citep{mushotzky00}, rather than the eight observed. 
The X-ray and visible-wavelength properties for our seven spectroscopic 
targets are provided in Table~\ref{tbl:obs}.

Of the six X-ray detected members of Abell 2104, only \#1 is a broad-line AGN 
and this is also the most luminous X-ray source. Only one of the remaining five 
has any emission lines: the weak [OIII] $\lambda5007$\AA\ in \#2 
(the atmospheric A band falls on H$\alpha$ at the cluster redshift). 
We did not detect [OII] $\lambda3727$\AA\, [OIII], or H$\beta$ emission 
in any of the remaining objects. 
The X-ray properties of the point sources were derived using the 
CIAO software package.
Sources \#1, 2, 5, and 7 were sufficiently bright 
to extract and fit X-ray spectra. All three of the cluster members are 
consistent with high absorbing columns and can only be AGN due to both their
luminosities and hard X-ray spectra.
We measured hardness ratios ($F_X {\rm [2-10keV]}/F_X {\rm [0.5-2keV]}$) 
for the remaining three members. 
Of these three, \#6 has a hardness ratio typical of a highly absorbed 
AGN. The remaining two (\#3 and 4) are softer, but still too hard to be
explained by thermal emission alone. The hardness ratios of these two objects
are consistent with a combination of thermal emission and a powerlaw
component. In this model, nearly all of the flux in the 2-10 keV band comes 
from the powerlaw component. Given the large luminosities of the powerlaw
components an AGN likely exists in both of these galaxies.

\section{Discussion}

Only one of the six AGN in Abell~2104, or $\sim 1$\% of the $\sim 100$ 
$R<20$ mag (rest-frame $M_V = -19.5$ mag) cluster members, is an obvious 
AGN in our optical spectra. 
This is in good agreement with the mean cluster AGN fraction of 1\% 
from optical spectroscopic surveys \citep{dressler99}. 
When \chandra\ observations are used to identify AGN, the AGN fraction 
appears to be a factor of five higher, or $\sim 5$\%. 
To evaluate the significance of this result, we use a simple binomial 
distribution test. Given an expected AGN fraction of 1\%, the variance in 
the number of AGN in a sample of 100 galaxies is $100 \times 0.01 \times 0.99$. 
We discovered five more AGN than expected, which is significant at the 
5/0.99 or $5\sigma$ level. 
In fact, an AGN fraction of 5\% is a lower limit to the true AGN fraction 
as our observations are not 
sensitive to more highly obscured AGN. For example, if the 
lower-luminosity AGN (such as \#5 and \#6) had a factor of five to ten higher 
absorbing columns, we would not have detected them; such higher absorption 
columns are not unusual among nearby field AGN with similar luminosities. 
There is additional evidence for a larger number of AGN in other clusters of 
galaxies in recent \chandra\ observations of point source excesses toward 
several other clusters \citep{cappi01,sun02,molnar02}.

Most of the AGN in Abell~2104 lack strong emission lines, which may be 
due to dilution by host galaxy light. For example, if we use the ratio of 
[OIII]$\lambda5007$\AA\ to hard X-ray flux for source \#1 to predict 
the equivalent widths for the other five sources, the equivalent widths would 
not be detectable in our spectroscopy. 
The [OIII]$\lambda5007$\AA\ to hard X-ray flux ratio for source \#1 is
$\sim 10^{-3}$ and is comparable to many Seyfert 2s \citep{polletta96}, 
although this ratio varies by several orders of magnitude in AGN. 
Our nondetection of emission lines in these objects could therefore be due to 
the low signal-to-noise of our spectroscopy, or due to some obscuration of 
the emission line region. 
It is not likely that the two X-ray sources with the lowest luminosities are 
powered by star formation, rather than AGN. While the most luminous local 
starbursts, such as Arp 220 or NGC 3256, have hard X-ray luminosities 
comparable to the two faintest sources in Abell 2104 \citep[$\sim 10^{41}$ 
\ergs][]{moran99,iwasawa99}, our low luminosity sources do not show detectable 
emission lines indicative of massive star formation. 
Furthermore, both Arp 220 and NGC 3256 would be 0.3 mag bluer in rest-frame 
$B-V$ (or 0.6 mag in observed $B-R$) than our A2104 optical counterparts. It is 
therefore more likely that these X-ray sources are powered by nuclear activity. 
Even if these two X-ray sources do not include an AGN component, the 
significance of our result changes to (4-1)/0.99 or a $3\sigma$ excess over 
expectations. 
These AGN, without detectable emission lines, appear similar to the optically 
bright, hard X-ray population that contribute approximately 40\% of the hard 
X-ray background \citep{mushotzky00}. 
If a significant number of such sources reside in clusters, they may be an 
important constituent of the hard X-ray background. 
Their association with the highly biased cluster galaxy population 
may also explain the variation in the hard X-ray background intensity 
from field to field discussed by \citet{cowie02}. 

The unexpectedly high AGN fraction in Abell 2104 is comparable to the fraction 
of galaxies that qualify as Butcher--Oemler galaxies \citep{butcher84} or that 
are emission-line galaxies \citep{liang00}. This AGN fraction, and the fact 
that it is a lower limit, suggests that obscured AGN may make a significant 
contribution to infrared and radio sources in cluster galaxies. 
For example, \citet{duc02} concluded that most of their FIR and radio-selected  
galaxies in Abell~1689 are powered by star formation because they only found 
optical emission line ratios consistent with AGN in one galaxy. 
Our result suggests that many of their sources could be AGN that lack strong 
emission-line features. 

Surprisingly, the host galaxies of these AGN all fall near or on the cluster 
color--magnitude relation \citep{sandage78}, which is presumably composed 
of old, quiescent galaxies without the reservoirs of cold gas necessary to 
fuel AGN. 
From our ground-based data ($1''$ seeing FWHM), approximately half of these 
host galaxies appear to have disk morphologies and half appear to have 
elliptical or S0 morphologies. They also span approximately three magnitudes 
in luminosity. 
Our naive expectation was that the AGN population in clusters should 
be associated with the blue, starforming galaxies as they have both 
significant reservoirs of cold gas and are interacting with the cluster 
potential for the first time. If the AGN lifetime were longer than the typical 
lifetime of $\sim 0.1$ Gyr for star forming galaxies in clusters 
\citep{poggianti99}, or there were a long delay before the onset of nuclear 
activity, AGN might instead be associated with the post-starburst population. 
In Abell~2104, the brightest three AGN shown in Figure~\ref{fig:cmd} fall 
on the cluster color--magnitude relation; while the fainter three are slightly 
bluer, they are not as blue as Butcher-Oemler galaxies. None of these AGN host 
galaxies have been completely stripped of their cold gas by the cluster 
potential \citep[\eg][]{gunn72}. 
These galaxies appear to have retained a reservoir of cold gas at the center 
of their potential wells similar to the dust disks observed in local cluster 
ellipticals \citep[\eg][]{jaffe94,martel99}. 

The association of these six AGN in Abell~2104 with red cluster galaxies could 
be because these observations only probe the inner 1~Mpc of the cluster, 
which is dominated by old, red galaxies. 
The population of blue, starforming galaxies in clusters is known to be 
more spatially extended \citep{butcher84} and have higher velocity dispersions 
\citep{dressler99} than the red, passively evolving galaxies. 
While the six AGN in Abell~2104 are photometrically and spectroscopically 
similar to old, red galaxies, their mean redshift is offset by $\sim 1000$ 
\kms\ from the cluster members measured by 
\citet{liang00}. At least some of the AGN may therefore be falling into the 
cluster for the first time. The asymmetric appearance of the cluster galaxy 
distribution in 
Figure~\ref{fig:image} suggests that this cluster is dynamically evolving. 
The spatial and kinematic distribution of AGN hosts to large cluster radii 
would help identify whether AGN are predominantly associated with the 
infalling galaxy population. 

The AGN host galaxy population in clusters could provide some interesting 
constraints on the fueling and lifetimes of AGN. Their association with blue 
galaxies would suggest that the AGN lifetime is less than the 0.1 Gyr lifetime 
of this population, consistent 
with current estimates for the lifetime of nearby AGN in the field 
\citep{martini02b}, and that nuclear activity is quenched by the cluster 
potential. 
If AGN are instead found in cluster galaxy populations of different ages, 
this suggests that nuclear activity is also episodic in the 
cluster environment and galaxies retain a central reservoir of cold gas for 
several Gyr after they enter the cluster potential. 
In either scenario, the presence of this large population of obscured AGN in 
cluster galaxies indicates that their central, supermassive black holes 
continue to grow in the cluster environment. 

\section{Summary}

We used deep \chandra\ and ground-based images to search for 
AGN in the cluster Abell~2104. Our follow-up spectroscopy has revealed 
six X-ray sources that are cluster members and all six fall on the cluster 
color--magnitude relation. 
The X-ray luminosities of four of these sources are sufficiently high that 
they can only be AGN. 
The remaining two are sufficiently luminous, red, and have hardness ratios 
that suggest they include or are dominated by an AGN component as well. 
These six sources correspond to an AGN fraction of $\sim 5$\% of cluster 
galaxies with $R<20$ mag. This is a lower 
limit to the cluster AGN fraction due to our insensitivity to AGN with 
higher absorbing columns. 
Only one of these six sources has the characteristic spectral signatures 
of AGN; the remaining five exhibit either very weak emission (one) or 
no evidence for nuclear activity (four). 
The fraction of optically identifiable AGN in Abell~2104 is therefore 
$\sim 1$\%, in good agreement with other clusters of galaxies 
\citep{dressler88,dressler99}. 

The X-ray selected AGN fraction is a factor of five greater than the 
fraction selected by optical spectroscopy alone and is comparable to the 
fraction of blue galaxies. 
The high fraction of X-ray selected AGN in Abell~2104 suggests the existence 
of an obscured, or at least optically unremarkable, AGN population in rich 
clusters of galaxies. Obscured AGN such as these could significantly 
affect infrared and radio-wavelength measurements of star formation in 
clusters of galaxies, as well as explain the cosmic variance in the hard 
X-ray background. Additional observations of AGN in other clusters of galaxies 
may provide interesting constraints on their fueling, lifetimes, and the growth 
of their central, supermassive black holes. 

\acknowledgements 

We would like to thank Alan Dressler, Chris Mihos, and Gus Oemler for helpful 
discussions and comments. We also acknowledge the referee for helpful comments 
that have improved our presentation. 
The {\it Chandra X-ray Observatory} Center is operated by the Smithsonian 
Astrophysical Observatory for NASA under contract NAS8-39073. 
PM was supported by a Carnegie Starr Fellowship. 
SCT was supported by NASA through Hubble Fellowship grant
HF-01125.01-99A awarded by the Space Telescope Science Institute,
which is operated by the Association of Universities for Research in
Astronomy, Inc., for NASA under contract NAS5-26555.  

%\bibliographystyle{apj}
%\bibliography{$HOME/tex/references/references.bib}

\clearpage

\setlength{\textwidth}{7.0in}
\setlength{\oddsidemargin}{-0.5in}
\setlength{\evensidemargin}{-0.5in}

\begin{deluxetable}{llcccccccl}
\tabletypesize{\scriptsize}
\tablecolumns{11}
\tablenum{1}
\tablewidth{550pt}
\tablecaption{Summary of the Observations\label{tbl:obs}}
\tablehead{
\colhead{ID} &
\colhead{Name} &
\colhead{$z$} &
\colhead{$N_H$} & 
\colhead{Index or $\Gamma$} & 
\colhead{$F_X$ [2-10keV]} & 
\colhead{$L_X$ [2-10keV]} & 
\colhead{$R$} & 
\colhead{$B-R$} & 
\colhead{Spectral Features} \\
\colhead{} &
\colhead{} &
\colhead{} &
\colhead{[$10^{22}$ cm$^{-2}$]} & 
\colhead{} & 
\colhead{[\ergs cm$^{-2}$]} & 
\colhead{[\ergs]} & 
\colhead{[mag]} & 
\colhead{} & 
\colhead{} \\
}
\startdata
1 & CXOU J154023.6-031347 & 0.159 & $1.99^{+0.20}_{-0.17}$    & $1.38^{+0.09}_{-0.08}$ & $1.62\times10^{-12}$   & $1.18\times10^{44}$ & 17.32 & 2.42 & [OIII],Mgb,NaD,H$\alpha$ \\ % 313
2 & CXOU J154016.7-031507 & 0.155 & $3.94^{+3.74}_{-2.07}$    & $1.33^{+1.31}_{-0.98}$ & $4.55\times10^{-14}$   & $3.59\times10^{42}$ & 18.79 & 2.16 & [OIII]$\lambda$5007,Mgb,NaD \\ % 571
3 & CXOU J154009.4-031519 & 0.162 &                           & 0.17                   & $2.8\times10^{-15}$    & $1.9\times10^{41}$  & 16.95 & 2.38 & H+K,G,Mgb \\ % 600
4 & CXOU J154014.0-031704 & 0.157 &                           & 0.16                   & $3.3\times10^{-15}$    & $2.2\times10^{41}$  & 19.67 & 2.04 & Mgb,NaD \\ % 917
5 & CXOU J154019.5-031825 & 0.161 & $0.35^{+0.23}_{-0.18}$    & $1.87^{+0.34}_{-0.30}$ & $1.80\times10^{-14}$   & $1.22\times10^{42}$ & 19.63 & 1.97 & G,Mgb \\ % 1177
6 & CXOU J154003.9-032039 & 0.154 &                           &                        & $1.34\times10^{-14}$   & $1.0\times10^{42}$  & 17.81 & 2.36 & Mgb,NaD \\ % 1436
7 & CXOU J154007.2-031952 & 1.51 & $0.087^{+0.043}_{-0.046}$ & $1.68^{+0.29}_{-0.27}$ & $8.43\times10^{-14}$   & $1.16\times10^{45}$ & 19.36 & 1.21 & MgII,OIII \\ % 1742
\enddata
\tablecomments{Summary of our observations of X-ray sources with bright, 
$R<20$ mag counterparts in the field of Abell 2104. 
The redshifts, best-fit absorbing column, spectral indices or hardness ratios, 
and observed hard X-ray fluxes are listed in columns 3 -- 6. We have used 
the spectral fits, or the models described below, to derive the 
absorption-corrected hard X-ray luminosities provided in column 7 
(assuming a ($\Omega_M,\Omega_{\Lambda},h) = (0.3,0.7,0.7)$ cosmology) . 
The errorbars are for the 90\% confidence level. 
Columns 8 \& 9 provide the $R$ magnitude and $B-R$ color of each source, 
and column 10 lists the spectral features used for the redshift determination. 
The fluxes for \#3 and \#4, which only have approximately ten counts each, 
were derived assuming a thermal model with a temperature of 1 keV and a power 
law with index 1.5. The flux for \#6 was derived assuming a power law with 
index 1.5 and an absorbing column $N_H = 1.75\times10^{22}$ cm$^{-2}$. These 
models are consistent with the measured hardness ratios. 
}
\end{deluxetable}

\end{document}